# DYNAMIC ENTROPIES, LONG–RANGE CORRELATIONS AND FLUCTUATIONS IN COMPLEX LINEAR STRUCTURES

WERNER EBELING, ALEXANDER NEIMAN and THORSTEN PÖSCHEL

*Humboldt–Universität zu Berlin, Institut für Physik,
Invalidenstr. 110, D–10115 Berlin, Germany*

We investigate symbolic sequences and in particular information carriers as e.g. books and DNA–strings. First the higher order Shannon entropies are calculated, a characteristic root law is detected. Then the algorithmic entropy is estimated by using Lempel–Ziv compression algorithms. In the third section the correlation function for distant letters, the low frequency Fourier spectrum and the characteristic scaling exponents are calculated. We show that all these measures are able to detect long–range correlations. However, as demonstrated by shuffling experiments, different measures operate on different length scales. The longest correlations found in our analysis comprise a few hundreds or thousands of letters and may be understood as long–wave fluctuations of the composition.

## 1 Introduction

The purpose of this paper is to compare different correlation measures based on methods of statistical physics. We aim to analyse correlations and fluctuations comprising at least several hundreds of letters. The characteristic quantities which measure long correlations are dynamic entropies [1,2,3], correlations functions and mean square deviations, $1/f^\delta$ noise [4,5], scaling exponents [6,7], higher order cumulants [8] and, mutual information [9,10]. Our working hypothesis which we formulated in earlier papers [1,5], is that texts and DNA show some structural analogies to strings generated by nonlinear processes at bifurcation points. This is demonstrated here first by the analysis of the behaviour of the higher order entropies. Further analysis is based on the mapping of the text to random walk models as well as on the spectral analysis. The random walk method which was proposed by Peng et al. [6], found several applications to DNA sequences [11,9] and to human writings [8,12]. In order to find out the origin of the long–range correlations we studied the effects of shuffling of long texts on different levels (letters, words, sentences, pages, chapters etc.). The shuffled sequences were always compared with the original one (without any shuffling). We note that all our files have the same letter distribution. However only the correlations on scales below the shuffling level are conserved. The correlations (fluctuations) on higher levels which are based on the large–scale structure of texts as well as on semantic relations are conserved only in the original file. The objects of our investigation were the book "Moby Dick" by Melville ($L \approx 1,170,000$ letters) and the DNA–sequence of the lambda–virus ($L \approx 50,000$ letters).

## 2 Entropy–like Measures of Sequence Structure

Let $A_1 A_2 \ldots A_n$ be the letters of a given substring of length $n \leq L$. Let further $p^{(n)}(A_1 \ldots A_n)$ be the probability to find in a string a block with the letters $A_1 \ldots A_n$. Then we may introduce the entropy per block of length $n$:

$$H_n = - \sum p^{(n)}(A_1 \ldots A_n) \log p^{(n)}(A_1 \ldots A_n) \tag{1}$$

$$h_n = H_{n+1} - H_n \tag{2}$$

Our methods for the analysis of the entropy of sequences were in detail explained elsewhere [13]. We have shown that at least in a reasonable approximation the scaling of the entropy against the word length is for large n given by root laws of the type

$$H_n = c_0 + c_1 \sqrt{n} + c_2 \cdot n \ . \tag{3}$$

$$h_n = 0.5\, c_1 \cdot n^{-\frac{1}{2}} + c_2 \ . \tag{4}$$

In our earlier work on Moby Dick [3] we assumed $c_2 = 0$ and obtained for our empirical data in the range $n = 10 - 26$ the constants $c_0 = 1.7$, $c_1 = 0.9$ . As shown already in 1951 by Shannon and a few years later by Burton and Licklider, the limit of the uncertainty $h_n$ for large n which is given here by $c_2$ should be finite. With the estimate $c_2 = 0.05$ we have repeated our fit including data for Moby Dick obtained with a new method [13]. The parameters obtained by the new fit are $c_0 = 1.7$, $c_1 = 0.5$, $c_2 = 0.05$. The dominating contribution ist given by the root term, an observation first made by Hilberg. In this way the decay of uncertainties $h_n$ follows a scaling according to a power law.

We consider now a different approach to the entropy analysis which goes back to the work of Kolmogorov, Chaitin, Lempel and Ziv. The algorithmic entropy according to Lempel and Ziv is introduced as the relation of the length of the compressed sequence (with respect to a Lempel–Ziv compression algorithm) to the original length.

The results obtained for the Lempel–Ziv complexities (entropies) of Moby Dick and of the lambda virus DNA are shown in Fig. 1. We see, that the Lempel–Ziv entropy of Moby Dick having a value of about 0.56 is much higher than than the Shannon entropies derived above. This clearly shows that the compression algorithms based on the Lempel-Ziv method are by far not optimal. As show our shuffling experiments represented in Fig. 1, compression algorithms of Lempel-Ziv type are mainly based on rules below the page level.

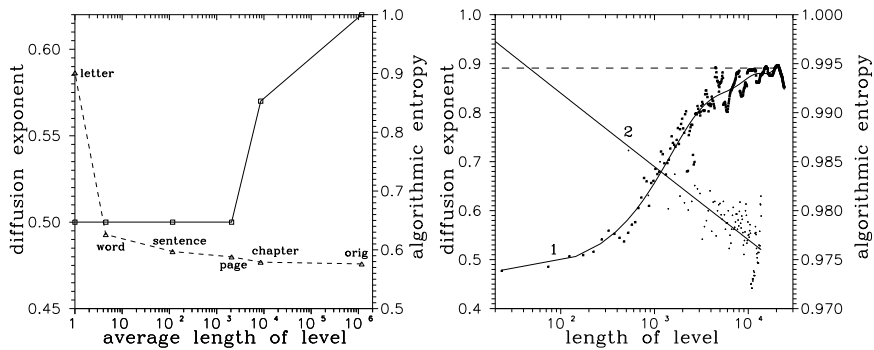

Figure 1: Lempel–Ziv complexities (dashed line) and scaling exponents of diffusion (full line) represented on the level of shuffling for the text Moby Dick (left) and for the Lambda–Virus DNA (right).

## 3  Correlation Functions, Power Spectra, Random Walk Exponents

In this part we closely follow the method proposed by Peng et al. [6,7] and the invariant representation proposed Voss [4]. Instead of the original string consisting of $\lambda$ different symbols we generate $\lambda$ strings on the binary alphabet (0,1) ($\lambda = 32$ for texts). In the first string we place a "1" on all positions where there is an "a" in the original string and a "0" on all other positions. The same procedure is carried also out for the remaining symbols. Then we generate random processes corresponding to these strings moving one step upwards for any "1" and remaining on the same level for any "0". The resulting move over a distance $l$ is called $y(k,l)$ where $k$ denotes the symbol. Then by defining a $\lambda$–dimensional vector space considering $y(k,l)$ as the component $k$ of the state vector at the (discrete) "time" $l$ we can map the text to a trajectory. The corresponding procedure is carried out for the DNA–sequences which are mapped to a random walk on a $\lambda = 4$–dimensional discrete space. The power spectrum is defined as the Fourier transform of the correlation function $C(k,n)$ which measures the correlation of the letters of type $k$ in a distance $n$ [5]. The results of spectra calculations for the original file of the Bible, for Moby Dick and for the same files shuffled on the word level or on the letter level correspondingly were presented in a foregoing work[8]. As a new result we present here the power spectrum of Moby Dick shuffled on the chapter and on the page level (Fig. 2).

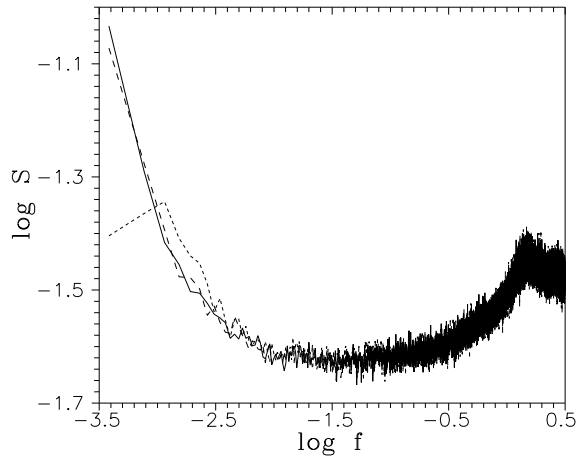

Figure 2: Double–logarithmic plot of the power spectrum for Moby Dick. The full line corresponds to the original text, the dashed line corresponds to the text shuffled on the chapter level and the dotted line to the text shuffled on the page level. Shuffling on a level below pages destroys the low frequency branch.

We see that the spectra of the original texts have a characteristic shape with a well–expressed low frequency part. This shows the existence of long–range correlations in Moby Dick. An estimate for the exponent is 0.8 . However it is difficult to extract a precise value for the exponent corresponding to a power law. One should better look at the shape of the whole curve which reminds a piecewise linear behaviour which is characteristic for multifractal structures. This point certainly needs further investigations. However we believe, there is no reason to expect that a long text as Moby Dick is a multifractal in a quantitative sense. On the other hand the hierarchical character of the structure of texts is beyond any doubt and we think that spectral curves may be characterised as a quantitative measure of this hierarchy. Fig. 2 shows that shuffling on the page level already destroys the low–frequency branch of the spectrum. This clearly proves that the origin of the $1/f$–fluctuations is on a scale which exceeds the page level. The contribution of high frequencies corresponds to the structure on the word and sentence level. Let us study now the anomalous diffusion coefficients which allow a higher accuracy of the analysis[7].

The mean square displacement for symbol $k$ is determined as

$$F^2(k,l) = <y^2(k,l)> - (<y(k,l)>)^2, \qquad (5)$$

where the brackets $< \cdot >$ mean the averaging over all initial positions. The behaviour of $F(k,l)$ for $l \gg 1$ is the focus of interest. It is expected that $F(k,l)$ follows a power law [7].

$$F(k,l) \propto l^{\alpha(k)}, \qquad (6)$$

where $\alpha(k)$ is the diffusion exponent for symbol $k$. We note that the diffusion exponent is related to the exponent of the power spectrum [7]. The case $\alpha(k) = 0.5$ corresponds to the normal diffusion or to the absence of long–range correlations. If $\alpha(k) > 0.5$ we have an anomalous diffusion which reflects the existence of long–range correlations. Beside the individual diffusion exponents for the letters we get also an averaged diffusion exponent $\alpha$ for the state space. The data are summarised in Fig. 1.

In the same way we can obtain other important statistical quantities: higher order moments and cumulants of $y(k,l)$ [8] By calculations of the Hölder exponents $D_q$ up to $q = 6$ we have shown, that the higher order moments show (in the limits of accuracy) the same scaling behaviour as the second moment. We repeated the procedure described above for the shuffled files. A graphical representation of the results for Moby Dick and DNA are given in Figs. 1.

We see from Figs 1–2 that the original sequences show strong long–range correlations, i.e. the coefficients of anomalous diffusion are clearly different from $1/2$ and there exists $1/f$–noise. After the shuffling below the page level the sequences become practically Bernoullian in comparison with the original ones since the diffusion coefficients decrease to a value of about $1/2$ and there is no more $1/f$–noise. The decrease occurs in the shuffling regime between the page level and the chapter level. For DNA–sequences the characteristic level of shuffling where the diffusion coefficient goes to $1/2$ is about 500–1000. Our result demonstrates that shuffling on the level of symbols, words, sentences or pages, or segments of length 500–1000 in the DNA–case destroys the long range correlations which are felt by the mean square deviations.

## 4 Conclusions

Our results show that the dynamic entropies, the low frequency spectra and the scaling of the mean square deviations, are appropriate measures for the long–range correlations in symbolic sequences. However, as demonstrated by shuffling experiments, different measures operate on different length scales. The longest correlations found in our analysis comprise a few hundreds or thousands of letters and may be understood as long–wave fluctuations of the composition. These correlations (fluctuations) give rise to the anomalous diffusion and to long–range $1/f$–fluctuations. These fluctuations comprise several hundreds or thousands of letters. There is some evidence that these correlations are based on the hierarchical organisation of the sequences and on the structural relations between the levels. In other words these correlations are connected with the grouping of the sentences into hierarchical structures as the paragraphs, the pages, the chapters etc. Usually inside certain substructure the text shows a greater uniformity on the letter level. In order to demonstrate this we have shown in earlier work [8] the (averaged over windows of length 4000) local frequency of the blanks (and other letters) in the text Moby Dick in dependence on the position along the text. The original text shows a large–scale structure extending over many windows. This reflects the fact that in some part of the texts we have many short words, e.g. in conversations (yielding the peaks of the space frequency), and in others we have more long words, e.g. in descriptions and in philosophical considerations (yielding the minima of the space frequency). The shuffled text shows a much weaker non–uniformity of the text, the lower the shuffling level, the larger is the uniformity. More uniformity means less fluctuations and more similarity to a Bernoulli sequence. For the case of DNA–sequences no analogies of pages, chapters etc. are known. Nevertheless the reaction on shuffling is similar to those of texts. Possibly a more careful comparison of the correlations in texts and in DNA sequences may contribute to a better understanding of the informational structure of DNA.